\documentclass[11pt]{article}
\usepackage[usenames,dvipsnames]{color}
\usepackage{hyperref}

\begin{document}
\title
{ Acoustic geometry through perturbation of mass accretion rate - radial 
flow in static spacetimes}
\author
{Deepika B Ananda{\footnote{anandadeepika[AT]hri.res.in}}\\
Indian Institute of Science Education and Research, Pashan, Pune-411008, INDIA\\
Harish-Chandra research Institute, Chhatnag Road, Jhunsi, Allahabad-211019, INDIA,~{\footnote{Current affiliation}}\\
Sourav Bhattacharya{\footnote{souravbhatta[AT]physics.uoc.gr}}
\\
Institute of Theoretical and Computational Physics, Department of Physics\\ University of Crete, 710 03 Heraklion, GREECE,\\
and\\
Tapas K Das~{\footnote{tapas[AT]hri.res.in}}\\
Harish-Chandra research Institute, Chhatnag Road, Jhunsi, Allahabad-211019, INDIA.
}

\maketitle
\abstract
In this work we present an alternative derivation of the general relativistic acoustic analogue geometry by perturbing the mass accretion rate or flux of an ideal fluid flowing radially in a general static and spherically symmetric spacetime. To the best of our knowledge, this has so far been done in non-relativistic scenario. The resulting causal structure of the two dimensional acoustic geometry is qualitatively similar to that one derives via the perturbation of the velocity potential. Using this, we then briefly discuss the stability issues by studying the wave configurations generated by the perturbation of the mass accretion rate, and formally demonstrate the stability of the accretion process.
This is in qualitative agreement with earlier results on stability, established via study of wave configurations generated by the perturbation of velocity potential, by using the acoustic geometry associated with it. We further discuss explicit examples of the Schwarzschild and Rindler spacetimes.
\vskip .5cm

{\bf Keywords:} {Analogue gravity, black hole accretion, stability
  

\vskip 1cm
\newpage
\maketitle
\section{Introduction}
When one considers propagation of an ideal irrotational fluid in a given background (flat or curved), it turns out that the potential corresponding to the irrotational fluid velocity propagates as a massless scalar field through an intrinsic curved manifold~\cite{M1980}-\cite{B1999} (see also~\cite{bar05, Barcelo:2004wz} for a comprehensive review). This internal effective manifold is known as the acoustic manifold and the corresponding metric is termed as the acoustic metric.

In this paper we consider the radial propagation of a general relativistic ideal fluid in a general static and spherically symmetric spacetime, and concern ourselves with the mass accretion rate, or the flux of mass of the fluid, and investigate whether the propagation of the linear
perturbation of the mass accretion rate, instead of the potential of the velocity, could be directly associated with any internal acoustic metric. Such connection indeed was earlier reported in~\cite{Naskar:2007su} for the Newtonian case. Here we wish to accomplish this task in a complete general relativistic framework.  
The chief motivation behind this study surely comes from the key role of the mass accretion rate in accretion astrophysics $-$ being a flux of mass of the fluid, this quantity can be observed and, in principle, can be measured using the present day
experimental techniques.

Investigation of the accretion process helps for the observational 
identification of the black hole candidates~\cite{SJS, FKR}. Stationary 
integral solutions of hydrodynamic transonic accretion onto astrophysical 
black holes are of great importance in identifying the observational 
signature of such compact objects through the construction of the 
emergent spectra. Since transient phenomena are, however, not quite
uncommon in large scale astrophysical set up, it is sometimes necessary 
to ensure that the stationary accretion solutions remain stable under 
perturbation, at least for astrophysically relevant time scales~\cite{garlick}-\cite{swagata}. 

To the best of our knowledge, the derivation of the acoustic geometry for irrotational flows and related stability issues 
for spherical accretion using it, was first done in~\cite{M1980}, using the linear perturbation of the velocity potential.


The present paper is our first work of the derivation of intrinsic acoustic geometry associated with the propagation 
of the perturbation of the mass accretion rate and related stability issues.
In this work we study the radial accretion flow in a general static and spherically symmetric spacetime. In our next set of works we plan to investigate the
axially symmetric matter flow for various representative black hole spacetimes - with 
and without the intrinsic spin angular momentum. Some 
preliminary results in that direction (axisymmetric flow in static spherically 
symmetric spacetime) has already been reported elsewhere~\cite{deepika-bose-institute-proceedings}. 
In recent years, connection between the stationary integral flow solutions describing the 
steady transonic 
black hole accretion and analogue effects has been established in~\cite{das04}-\cite{pratik}.

Then it is natural to ask, can we establish such connections via perturbation analysis? 
How does the acoustic metric differ
from that one obtains via the well known perturbation of the velocity potential? 
Also, can we establish the stability issues using this approach? 
In particular, we show that the perturbation indeed propagates through
an effective $(1+1)$-dimensional geometry, which comprises of the time and radial directions.
We note that there are some ambiguities in defining an acoustic metric in $(1+1)$-dimensions via the velocity
potential approach, as one 
encounters an overall conformal factor which is divergent~\cite{bar05, Barcelo:2004wz}. There are
two possible ways to deal with this problem. Firstly, to take into account the extra dimensions and make
the metric three or more dimensional. The second way is to consider the `densitied' symmetric
tensor itself, which appears naturally in the perturbation equation. One may regard this tensor as an `effective' acoustic metric
as this provides the acoustic geometrical structure anyway. We shall see that we encounter same difficulty in dealing with the mass accretion rate as well in Section 3. However, for the densitied effective metric which is finite for both these approaches, we find a mismatch between an overall conformal factor. This difference can be attributed to the
fact that the mass accretion rate and the potential corresponding to the velocity are two different 
physical entities, from the qualitative as well as the quantitative viewpoint. We shall be more explicit on this in the following
sections.

Nevertheless, we demonstrate that the causal structure of the acoustic geometry remains the 
same irrespective of whether they are obtained via perturbation of the velocity potential or the 
mass accretion rate. This conclusion holds, as we will show in our forthcoming papers, 
for axisymmetric flow as well. This connection between the propagation of the mass accretion rate and the acoustic causal structure
seems to be interesting in its own right. Moreover, we study various wave configurations generated by the perturbation of the mass accretion rate and argue the stability of the accretion process. This is in qualitative agreement with the earlier result of~\cite{M1980}, established via the study of wave configurations generated by the perturbation of the velocity potential. The approach
of~\cite{M1980} seems to be extremely useful in constructing in particular, energy integrals and its time derivative. On the other hand, apart from the demonstration of the existence of the acoustic causal structure associated with the perturbation of the mass accretion rate itself, our approach seems to be convenient in getting an estimate of the profile of the perturbed mass accretion rate, which is nevertheless also a physically meaningful quantity in accretion astrophysics.


In subsequent sections, we formulate the linear perturbation analysis scheme to 
derive the acoustic geometry for a general static and spherically symmetric spacetime
for radial flow and also briefly address the stability issues. Based on such general 
formalism we then demonstrate some specific results for the Schwarzschild and the Rindler spacetime.
For the Rindler spacetime we show that, if the velocity of the fluid is smooth, there can be no sonic 
horizon and we discuss its connection with the Minkowski spacetime. 

We shall use mostly positive signature for the metric, and will set $c=1=G$ throughout.

\section{The basic constructions}
Let us begin by considering an ansatz for the metric describing a general static and spherically symmetric spacetime
\begin{equation}
ds^{2}= -g_{tt}(r) dt^{2}+g_{rr}(r)dr^{2}+ r^2 \left(d\theta^{2}+\sin^2\theta d\phi^{2}\right),
\label{eqn4.1}
\end{equation}
where we do not necessarily have $ g_{tt}g_{rr}=1$. For a static and spherically symmetric spacetime, $g_{tt}g_{rr}\neq 1$, when the energy-momentum tensor does not satisfy $T^{\rm B}_t{}^{t}+T^{\rm B}_r{}^r= 0$~\cite{Jacobson:2007tj}, where $T^{\rm B}_{\mu\nu}$ corresponds to some background matter field which backreacts (i.e., not of the test fluid we are studying).

We consider a perfect fluid with the energy-momentum tensor
\begin{equation}
T_{\mu \nu} = \left(\epsilon + p \right) v_{\mu} v_{\nu} + pg_{\mu \nu},
\label{eqn4.4}
\end{equation} 
where $\epsilon$ is the energy density, $p$ is the pressure and $v^{\mu}$ is the fluid's four-velocity, normalized as $v^{\mu}v_{\mu}=-1$. We assume the only non-vanishing spatial part of $v^{\mu}$ to be the radial component, $v$. Then the normalization condition yields
\begin{equation}
v^{t}= \sqrt{\frac{1+g_{rr} v^{2}}{g_{tt}}}.
\label{eqn4.2}
\end{equation}
We shall ignore throughout the backreaction of the fluid on the spacetime metric.

We assume that the fluid satisfies the equation of state of an ideal gas under the barotropic 
conditions, and is described by adiabatic equation of state, $p = k \rho^{\gamma}$, where $\rho$ 
is the fluid's number density and $\gamma$ is the polytropic index. 
The specific enthalpy is defined as $h = \frac{\epsilon+p}{\rho}$, so that
\begin{equation}
dh = T d \left( \frac{S}{\rho} \right) + \frac{dp}{\rho},
\label{tr}
\end{equation}
where $T$ and $S$ are respectively the temperature and the entropy of the fluid. The adiabatic sound speed is defined under the isoentropic condition as
\begin{equation}
c_{\rm s}^2 = \frac{\partial p}{\partial \epsilon}\Bigg\vert_{dS=0}.
\end{equation}  
The continuity equation ensures conservation of mass, $\nabla_{\mu} \left(\rho v^{\mu} \right)=0$. 
Since we are interested only in the radial motion, we have 
\begin{eqnarray}
\frac{1}{\sqrt{-g}}\partial_{t} \left( \rho v^{t} \sqrt{-g} \right) + \frac{1}{\sqrt{-g}}\partial_{r} \left( \rho v \sqrt{-g} \right) =0.
\label{eqna1}
\end{eqnarray} 
Also, the conservation equation for the energy-momentum tensor is
\begin{equation}
\nabla_{\mu} T^{\mu \nu} = 0.
\label{eqna2}
\end{equation}
With the help of the continuity equation and the thermodynamic relations stated above, Eq.~(\ref{eqna2}) can be recast as
\begin{equation}
v^{\nu} \partial_{\nu} v^{\mu} + \Gamma^{\mu}_{\nu \lambda} v^{\nu} v^{\lambda} + \frac{c_{\rm s}^{2}}{\rho} \left( v^{\mu} v^{\nu}  + g^{\mu \nu} \right) \partial_{\nu} \rho = 0.
\label{eqna6}
\end{equation}
As both $v^{\theta}$ and $v^{\phi}$ vanish, the only relevant Christoffel connections are
\begin{eqnarray}
\Gamma^{t}_{tr} = \frac{1}{2} g^{tt} \partial_{r} g_{tt},~
\Gamma^{r}_{tt} = \frac{1}{2}g^{rr} \partial_{r}g_{tt},~
\Gamma^{r}_{rr} = \frac{1}{2}g^{rr} \partial_{r}g_{rr}.\nonumber \\
\label{eqna9}
\end{eqnarray}
For $\mu=t$, we get the energy conservation equation 
\begin{equation}
v^{t} \partial_{t} v^{t} + v \partial_{r} v^{t} + g^{tt} \left( \partial_{r} g_{tt} \right)v v^{t} + \frac{c_{\rm s}^{2}}{\rho} \left[ \left( (v^{t})^{2}- g^{tt} \right)\partial_ {t} \rho + v v^{t} \partial_ {r} \rho\right]  = 0. 
\label{eqna8}
\end{equation}
Similarly, $\mu=r$ gives the radial Euler equation which ensures momentum conservation along radial direction 
\begin{equation}
v^{t} \partial_{t} v+ v \partial_{r} v + \frac{(v^{t})^{2}\partial_{r}g_{tt}}{2g_{rr}} +  \frac{v^{2}\partial_{r}g_{rr}}{2g_{rr}}+ \frac{c_{\rm s}^{2}}{\rho} vv^{t} \partial_{t} \rho + \frac{c_{\rm s}^{2}}{\rho} \left[ v^{2} + g^{rr} \right] \partial_{r} \rho = 0.
\label{eqna10}
\end{equation}
For convenience, using the normalization condition we rearrange the above equation to get
\begin{equation}
\sqrt{\frac{1+g_{rr} v^{2}}{g_{tt}}} \partial_{t} v + v \sqrt{\frac{1+g_{rr} v^{2}}{g_{tt}}} \frac{c_{\rm s}^{2}}{\rho} \partial_{t} \rho + \partial_{r} \left(\frac{1+g_{rr} v^{2}}{2g_{rr}}\right) + \left(\frac{1+g_{rr} v^{2}}{2g_{rr}}\right) \frac{\partial_{r} (g_{tt}g_{rr})}{g_{tt}g_{rr}} + \frac{c_{\rm s}^{2}}{\rho} \left(\frac{1+g_{rr} v^{2}}{g_{rr}} \right) \partial_{r} \rho = 0.
\label{eqna12}
\end{equation}
Let us now briefly describe for our future convenience the stationary solutions for the accretion process in the general static and spherically symmetric metric we use. 

Since the spacetime is spherically symmetric, we take the density and velocity components to be independent of $\theta$  and $\phi$. For the stationary state the continuity equation (Eq.~(\ref{eqna1})) becomes $\partial_{r}\left(\rho_0 v_0 \sqrt{-\widetilde{g}}\right)=0$, where
\begin{equation}
 \widetilde{g}=\frac{g}{\sin^2\theta}= -g_{tt}g_{rr}r^4,
\label{det}
\end{equation}
and hereafter the subscript `$0$' will correspond to the stationary values. Integrating this we obtain $ \rho_0 v_0 \sqrt{-\widetilde{g}}={\rm constant}$. We multiply both sides of this equation with the element of solid angle $d\Omega=\sin^2\theta d\theta d\phi$
and integrate to find
\begin{equation}
\Omega \rho_{0} v_{0} \sqrt{-\widetilde{g}} = - \dot{M} = {\rm constant},
\label{eqn4.9}
\end{equation}
where $\Omega$ is the solid angle that appears due to angular integration over the 
surface through which the mass flux is being estimated. For radial flow from all directions, clearly one has $\Omega=4\pi$. 
 The constant $\dot{M}$
is called the mass accretion rate and the negative sign conventionally signifies the infall of the matter. 
Since $\Omega$ is a constant, it can be absorbed on the right hand side of Eq.~(\ref{eqn4.9}) to redefine the mass accretion rate. 
So without any loss of generality, henceforth we shall suppress $\Omega$ and simply express
the mass accretion rate as $\rho_{0} v_{0} \sqrt{-\widetilde{g}}= \Psi_0~(\rm say)$.

Integration of the time independent part of Eq.~(\ref{eqna8}) provides
\begin{equation}
h v_{t} = -\cal{E},
\label{eqn4.10} 
\end{equation}
where ${\cal E}$ is a constant along the streamlines and is identified as the specific energy (or, the 
relativistic Bernoulli's constant). Following similar procedure  Eq.~(\ref{eqna12}) gives
\begin{equation}
-\frac{2c_{\rm s_{0}}^{2}}{\rho_{0}}\partial_{r} \rho_{0} = \left(\frac{g_{rr}}{1+g_{rr} v_{0}^{2}}\right) \partial_{r} \left(\frac{1+g_{rr} v_{0}^{2}}{g_{rr}}\right) +  \frac{\partial_{r} (g_{tt}g_{rr})}{g_{tt}g_{rr}}.
\label{eqn4.11}
\end{equation}
For convenience it is customary to move to a local Lorentz-like frame. Denoting the radial velocity by $u$ in such frame, we define
\begin{eqnarray}
v^{t} = \frac{1}{\sqrt{g_{tt}}} \frac{1}{\sqrt{1-u^{2}}},\quad
v = u \sqrt{\frac{1}{g_{rr}\left(1-u^{2}\right)}}.  
\label{eqs1}
\end{eqnarray}
By substituting the above into Eq.s~(\ref{eqn4.9}), (\ref{eqn4.11}), stationary solutions which were initially a function of $v$ can now be expressed in terms of $u$. Considering the radial derivative of Eq.~(\ref{eqn4.9}) and
using it into Eq.~(\ref{eqn4.11}), we find
\begin{equation}
\frac{du_0}{dr} = \frac{u_{0}\left(1-u_{0}^{2}\right)\left[c_{\rm s_{0}}^{2}\partial_{r} \ln \left(\frac{-\widetilde{g} }{g_{rr}}\right)-\partial_{r}\ln(g_{tt})\right]}{2\left( u_{0}^{2}-c_{\rm s_{0}}^{2}\right)}.
\label{eqs3}
\end{equation}
The denominator vanishes when $u_{0}^{2} = c_{\rm s_{0}}^{2}$,
which implies that the accretion solutions are transonic. The flow encounters a sonic point, say at $r_{\rm h}$, at which the speed of the flow is equal to the sound speed. To make $\frac{du_0}{dr}$ finite at the sonic point, we simultaneously set the numerator to zero as well, which gives the velocity or the sound speed there, 
\begin{equation}
u_{0}\vert_{h}=c_{\rm s_{0}} \vert_{h} = \frac{\partial_{r} \ln \left(g_{tt} \right)}{\partial_{r}\ln \left(-\widetilde{g} \right) - \partial_{r} \ln g_{rr}},
\label{uh}
\end{equation}
where the subscript $h$ implies that corresponding quantities are evaluated at the sonic point. A detailed discussion of the above things for the Schwarzschild spacetime can be 
found in~\cite{das04}\footnote{The steady transonic 
Bondi solutions in the Schwarzschild metric has also been 
discussed in~\cite{michel,begelman,malec}, from astrophysical 
perspective.}.
With all these ingredients and formulae, we are now ready to show the emergence of acoustic geometry via the perturbation of the mass accretion rate.

\section{The emergence of acoustic geometry and related stability issues }
\subsection{The geometry}
Let us first see the geometric properties of the mass accretion rate. Since $\Psi_0=\rho_{0} v_{0} \sqrt{-\widetilde{g}}$, where $\rho_0$ is a scalar, $v_0$ is the component of velocity and $\widetilde{g}$ is related to the determinant of the metric (Eq.~(\ref{det})), it is clear that the mass accretion rate cannot behave like a scalar under coordinate transformations, unlike the velocity potential. 

By defining the velocity potential $\Phi$, $hv_{\mu}=\nabla_{\mu}\Phi$ for an irrotational flow, we have
 \begin{eqnarray}
\Psi_0=\frac{\rho_0\sqrt{-\widetilde{g}} \left(\partial_r \Phi \right)}{h},
\label{g1}
\end{eqnarray}
which obviously does not behave as a scalar. In fact any flux depends upon component of velocity and some surface and hence can in general  not expected to be a scalar quantity. This is clearly a crucial qualitative difference between the velocity potential and the mass accretion rate. 

 However, since we are dealing with static and spherically symmetric spacetimes, the coordinate system of Eq.~(\ref{eqn4.1}) is a natural choice with respect to all the four isometries (three spatial rotations and one time translation) of the spacetime. Thus we are working in an isometry invariant framework, just like Lorentz invariance in flat spacetime's particle physics.  
In any case, we shall see below that the acoustic causality the mass accretion rate is associated with, has a covariant meaning.

Let us employ a linear perturbation scheme,
\begin{eqnarray}
v(r,t) = v_{0}(r)+v'(r,t),\\
\rho(r,t) = \rho_{0}(r)+\rho'(r,t).
\label{eqn4.12}
\end{eqnarray}
Also, we define a variable $\Psi=\rho v \sqrt{-\widetilde{g}}$, so that its stationary value corresponds to $\Psi_{0}$, defined in the previous section.
We thus have up to linear terms,
\begin{eqnarray}
\Psi'(r,t) &=& \left( \rho'v_{0}+\rho_{0}v' \right)\sqrt{-\widetilde{g}}.
\label{eqn4.14}
\end{eqnarray}
Substituting the perturbed quantities into Eq.~(\ref{eqna1}), we have
\begin{equation}
\partial_{t} \left[ \rho'v_{0}^{t} + \frac{g_{rr} \rho_{0} v_{0} v'}{g_{tt}v_{0}^{t}} \right] + \frac{1}{\sqrt{-\widetilde{g}}}\partial_{r} \Psi' = 0,
\label{eqn4.15}
\end{equation}
where $v_{0}^{t} = \sqrt{\frac{1+g_{rr} v_{0}^{2}}{g_{tt}}}$. Using Eq.s~(\ref{eqn4.14}) and (\ref{eqn4.15}), one can express the derivatives of $\rho'$ and $v'$ solely in terms of $\Psi'$,
\begin{eqnarray}
\partial_{t} v' = \frac{g_{tt} v_{0}^{t}}{\rho_{0}\sqrt{-\widetilde{g}}}  \left[v_{0}^{t} \partial_{t} \Psi'+v_{0} \partial_{r} \Psi' \right],
\label{eqn4.17} \\
\partial_{t} \rho' = -\frac{1}{\sqrt{-\widetilde{g}}} \left[ g_{rr}v_{0} \partial_{t} \Psi' + g_{tt} v_{0}^{t}  \partial_{r} \Psi' \right].
\label{eqn4.18}
\end{eqnarray}
Substituting for the perturbed quantities in the radial equation, Eq.~(\ref{eqna12}) and using Eq.~(\ref{eqn4.11}) gives
\begin{equation}
\frac{g_{rr}}{g_{tt} v_{0}^{t}}\partial_{t} v' + \partial_{r} \left[ \frac{g_{rr} v_{0}v'}{g_{tt} (v_{0}^{t})^{2}}+ \frac{c_{\rm s_{0}}^{2}\rho'}{\rho_0} \right] + \frac{g_{rr} v_{0} c_{\rm s_{0}}^{2}}{g_{tt} \rho_{0}v_{0}^{t}} \partial_{t} \rho' = 0.
\label{eqn4.19}
\end{equation}
Taking the time derivative of the above equation and substituting for the time derivatives of $\rho'$ and $v'$ from Eq.s~(\ref{eqn4.17}), (\ref{eqn4.18}), we obtain
\begin{eqnarray}
\partial_{t}\left( \frac{g_{rr} v_{0}}{v_{0}^{t}} \left[ \frac{c_{\rm s_{0}}^{2}+\left( 1- c_{\rm s_{0}}^{2} \right)g_{tt}(v_{0}^{t})^{2}}{g_{tt}} \right] \partial_{t} \Psi'\right)+ \partial_{t}\left( \frac{g_{rr} v_{0}}{v_{0}^{t}}\left[v_{0}v_{0}^{t}\left(1-c_{\rm s_{0}}^{2} \right)\right] \partial_{r} \Psi'\right)+
\nonumber \\
\partial_{r}\left( \frac{g_{rr} v_{0}}{v_{0}^{t}}\left[v_{0}v_{0}^{t}\left(1-c_{\rm s_{0}}^{2}\right)\right] \partial_{t} \Psi'\right)+\partial_{r}\left( \frac{ v_{0}}{v_{0}^{t}}\left[-c_{\rm s_{0}}^{2} + \left(1-c_{\rm s_{0}}^{2} \right) g_{rr} v_{0}^{2} \right] \partial_{r} \Psi'\right)=0,
\label{eqn4.20a}
\end{eqnarray}
so that now we can identify a symmetric $2\times2$ matrix $f^{\mu\nu}$ {\it through which} the perturbation $\Psi'(r,t)$ of the mass accretion rate propagates, 
 \[ 
f^{\mu \nu} \equiv 
\frac{g_{rr} v_{0} c_{\rm s_{0}}^{2}}{v_{0}^{t}} \left[ \begin{array}{ccc}
-g^ {tt} + \left(1- \frac{1}{c_{\rm s_{0}}^{2}}\right)(v_{0}^{t})^{2} &  v_{0}v_{0}^{t} \left(1-\frac{1}{c_{\rm s_{0}}^{2}}\right) \\ 
\\ v_{0}v_{0}^{t}\left(1-\frac{1}{c_{\rm s_{0}}^{2}}\right) & g^{rr} + \left(1-\frac{1}{c_{\rm s_{0}}^{2}} \right) v_{0}^{2}\\
 \end{array} \right], \]
\label{metric}
such that 
\begin{equation}
\partial_{ \mu} \left( f^{\mu \nu} \partial_{ \nu} \Psi' \right) = 0.
\label{eqn4.20b}
\end{equation}
It is clear that $f^{\mu \nu}$ can be formally expressed as 
\begin{eqnarray}
f^{\mu\nu} = \zeta^{2}\left[g^{\mu\nu}+\left(1-c^{-2}_{\rm s_0}\right)v_0^{\mu}v_0^{\nu}  \right],
\label{am1}
\end{eqnarray}
with $\zeta^2=\frac{g_{rr} v_{0} c_{\rm s_{0}}^{2}}{v_{0}^{t}}$. We note here that the part of $f^{\mu\nu}$ within square bracket
is a Rank 2 tensor and the {\it same} as one gets via the well known perturbation of the velocity potential~\cite{M1980}-\cite{B1999}. 
The overall factor $\zeta^2$ is clearly not a scalar, due to the appearances of the metric and velocity components. Also, the apparent mismatch 
in the sign before the second term with~\cite{B1999} is due to the fact that we are working in a mostly positive signature of the metric for which $v^{\mu}v_{\mu}=-1$ instead of $+1$.

The non-scalar nature of $\zeta^2$ should be related to the non-scalar nature of the mass accretion rate itself, as discussed at the beginning of Section 3.

The inverse $f_{\mu\nu}$ of  $f^{\mu\nu}$ can be defined as
\begin{eqnarray}
f_{\mu\nu} = \zeta^{-2}\left[g_{\mu\nu}+\left(1-c^{2}_{\rm s_0}\right)v_{0\mu}v_{0\nu}  \right],
\label{am2}
\end{eqnarray}
so that $f^{\mu\nu}f_{\mu\lambda}=\delta^{\nu}{}_{\lambda}$.

We note that except on a black hole horizon, the metric function $g_{rr}$ is expected to be well behaved. Thus, the acoustic causal structure described by $f_{\mu\nu}$ is qualitatively the same as one gets via the perturbation of the velocity potential.
In particular, we note that on or in the infinitesimal vicinity of a black hole horizon, the linear perturbation scheme we are using may not be very meaningful.

The nonrelativistic limit of $f_{\mu\nu}$ can be achieved by taking $\left\vert g_{\mu\nu}\right\vert\to 1$
and $v_0,c_{\rm s_0}\ll1$,
\[ 
f_{\mu \nu}\Big\vert_{\rm NR} \equiv 
\frac{1}{v_0 c^2_{\rm s_{0}}} \left[ \begin{array}{ccc}
 v_{0}^{2} - c_{\rm s_{0}}^{2} & -v_{0}  \\ 
\\ -v_{0} & 1\\
 \end{array} \right],\]
derived first in~\cite{unr81}, also later in~\cite{Naskar:2007su}.

We may proceed to define the so called acoustic metric from $f^{\mu\nu}$, following exactly the same way as is done in
the velocity potential perturbation. Precisely, one writes
\begin{equation}
\sqrt{-G}G^{\mu \nu} = f^{\mu \nu},
\label{eqn4.22}
\end{equation}
for some $G^{\mu\nu}$ and $G$ is the determinant of the inverse of $G^{\mu\nu}$. Then one finds $G_{\mu\nu}$, which is regarded as the acoustic metric. But as we mentioned earlier, this procedure leads to a divergent conformal factor, and hence renders the acoustic metric ill defined in $(1+1)$ dimensions~\cite{bar05, Barcelo:2004wz}. 

Let us see this a bit more explicitly, the essential details of which can be found in~\cite{Barcelo:2004wz} (see also~\cite{bar05}).
Let us imagine that Eq.~(\ref{eqn4.22}) holds in some arbitrary dimension $n$, for some well behaved $f^{\mu\nu}$. Then by taking the determinant of Eq.~(\ref{eqn4.22}) we get 
\begin{equation}
|\det G|=|\det f^{\mu\nu}|^{\frac{2}{n-2}},\quad{\rm and}\quad G^{\mu \nu} = |\det f^{\mu \nu}|^{-\frac{1}{n-2}}f^{\mu\nu},
\label{eqn4.22ad1}
\end{equation}
which immediately shows the problem of working in $n=2$, due to appearances of diverging or vanishing factors associated with the determinant -- we have $G^{\mu\nu}=0$ for a well behaved $f^{\mu\nu}$. 
 
However, as has been argued in~~\cite{Barcelo:2004wz}, this issue seems to be more formal than fundamental. One can easily deal with this 
by adding extra dimensions to make $n>2$, or can just work with $f^{\mu\nu}$ only. After all, since $f_{\mu\nu}$ is well behaved,
the behaviour of $\Psi'$ (or, for the standard approach, the behaviour of the perturbation of the velocity potential) should entirely be determined with the help of it and some suitable boundary conditions. Putting these all in together, we shall proceed with $f^{\mu\nu}$ or $f_{\mu\nu}$, and just call them effective acoustic metric, without compromising any physical content of the accretion phenomenon. 

We note that the acoustic geometry derived in~\cite{M1980} for spherical accretion in the Schwarzschild spacetime is of dimension four, and $G_{\mu\nu}$ shows no problem at all when we go to dimension two by setting
`$\theta$' and `$\phi$' to constants. The reason is, for any dimension greater than two, the determinants in Eq.s~(\ref{eqn4.22ad1}) are well behaved. And setting some coordinates to constants still keeps those factors well behaved (as we still have $n=4$ in the multiplicative determinant functions), while reducing
the number of the degrees of freedom of the tensorial part (i.e. non-determinant part). This is equivalent to projecting a given well behaved $4\times 4$ matrix onto a $2\times 2$ subspace.

Clearly, this has no contradiction with our result as we have worked throughout in dimension two here. In other words, there should be some qualitative difference between, while we are deriving throughout something in dimension $n$, and going to dimension $n$ from some higher dimensional theory. For the present case Eq.s~(\ref{eqn4.22ad1}) manifest that.

A very simple example of this qualitative mismatch can be given from the Schwarzschild spacetime written in spherical coordinates, for which $|\det g|=r^4\sin^2\theta$. On any $\theta={\rm const.}=\theta_{0}$ and $\phi={\rm const.}$ plane, it has value $r^4\sin^2\theta_0$. On the other hand, if we consider $1+1$-dimensional Schwarzschild spacetime from the very beginning in the `$t-r$' plane, we always have $|\det g|=1$.

Finally we also note that since $\Psi'$ is not a scalar, attempting to write down Eq.~(\ref{eqn4.20b}) in a form like 
$\frac{1}{\sqrt{-G} }\partial_{\mu}\left(\sqrt{-G} G^{\mu\nu}\partial_{\nu} \Psi'\right)=0$, even in dimension higher than two,
has no clear geometrical advantage.

One remaining crucial question now is : is the acoustic geometric causal structure, i.e. the sonic horizon, we have found in $f_{\mu\nu}$ has some covariant meaning? The answer is Yes. This is just because the essential information about the sonic horizon in $f_{\mu\nu}$ (Eq.~(\ref{am2})) is contained within its tensorial part.


\vskip .2cm

To summarize, our analysis shows that the acoustic geometry is qualitatively independent of the two ways of perturbation. For nonrelativistic case this equality was demonstrated earlier in~\cite{Naskar:2007su}, and we have demonstrated here that this holds true for general relativistic case as well. The velocity potential and the mass accretion rate are apparently two very distinct quantities. In fact if we rewrite  Eq.~(\ref{eqn4.20b})
in terms of the perturbation of the velocity potential, say $\Phi'$ using Eq.~(\ref{g1}), we get a third order differential equation for $\Phi'$.
Thus the result we have obtained here is far from obvious {\it a priori}. 

We can study the acoustic geometrical properties by defining notions of acoustic event horizon, acoustic apparent horizon, ergosphere, surface gravity and so on. We shall not go into details of this here, referring our reader to e.g.~\cite{B1999} for this.

Having found the perturbation equations and the acoustic geometry through which the perturbation of the mass accretion rate propagates, it is worthwhile to briefly address the stability issues, which are of chief importance in astrophysical scenarios. 

The stability issues using acoustic metric obtained via the perturbation of the velocity potential 
has been first discussed extensively in~\cite{M1980} for spherical Bondi accretion onto a Schwarzschild black hole.
Using our alternative effective metric $f_{\mu\nu}$ for a general static and spherically symmetric spacetime, we shall demonstrate briefly below 
how we reestablish those stability issues.  

Here we note the crucial difference between the approaches of stability analysis of the following and that of~\cite{M1980}. Precisely,
we shall study the possible standing and travelling wave configurations generated by the perturbation of the mass accretion rate, using  
Eq.~(\ref{eqn4.20b}). Ref.~\cite{M1980}, on the other hand, studies such waves generated by the perturbation of the velocity potential using the acoustic geometry associated with it and also constructs energy integrals. 

In any case, we shall see below that the geometry we have found ensures stability for such wave configurations generated by the perturbation of the mass accretion rate, and hence in qualitative agreement with the earlier results.

\subsection{The stability issues }
Let us start studying the stability analysis by assuming a trial wave solution for the perturbation of the mass accretion rate,
\begin{equation}
\Psi'_{\omega}(t,r) = p_{\omega}(r) \exp(-i\omega t),
\label{s2}
\end{equation}  
and substitute into Eq.~(\ref{eqn4.20b}) to get
\begin{equation}
\omega^{2} p_{\omega}(r) f^{tt} + i \omega \left[\partial_{r} \left( p_{\omega}(r)f^{rt} \right)+ f^{tr} \partial_{r} p_{\omega}(r)]-[\partial_{r} \left(f^{rr} \partial_{r} p_{\omega}(r) \right) \right] = 0. 
\label{eqn4.52}
\end{equation}
It is reasonable to assume that the flow is subsonic at large distances from the gravitating source. Now, we have two kind of solutions possible for fluid accreting onto a gravitating source -- subsonic and transonic solutions (see e.g.~\cite{PSO1980} for details). The subsonic solutions work perfectly well for compact objects with a hard surface like neutron star but not for a black hole. The second kind, transonic Bondi solutions~\cite{Bondi1952}, goes through a sonic point where the flow velocity equals the sound speed and at the smaller radii the flow becomes supersonic. This cannot be smoothly matched on to a hard stellar surface for obvious reasons that the flow cannot be supersonic there. Though one may try to address this issue through a shock, it is physically not possible to incorporate shock here, as presently we are dealing with adiabatic equation of state. Thus for our present case the transonic Bondi solutions are meant for black holes only.

Now, the standing waves require the perturbation to vanish at the boundaries at all times. The outer boundary can be at very large radius, for example, at the capture radius of the accretor. The inner boundary can be the surface of the object itself. If the compact object is a black hole, then the unperturbed flow is described by the Bondi solutions and there is no physical mechanism to constrain the flow in the supersonic region to have vanishing perturbations, which is mandatory to have standing waves. 
Since the standing wave analysis rely on the continuity of the solution, we thus have to restrict ourselves to completely subsonic flows.

Hence we choose two boundaries at $r_{1}$ and $r_{2}$ such that $p_{\omega}(r_{1}) = 0=p_{\omega}(r_{2})$, and we consider the properties of standing wave in the region between these boundaries, $r_{1} \leq r \leq r_{2}$. Multiplying Eq.~(\ref{eqn4.52}) with $p_{\omega}(r)$ and integrating by parts we find
\begin{equation}
\omega^{2}= - \frac{\int f^{rr} \left( \partial_{r} p_{\omega}(r) \right)^{2}}{\int f^{tt} p^2_{\omega}(r)}.
\label{eqn4.29a}
\end{equation}
It is easy to see that the denominator stays negative due to $f^{tt}$. So the solutions for 
$\omega$ depend only on the sign of $f^{rr}$. For subsonic flows, $\frac{g_{rr}v_{0}^{2}}{1+g_{rr}v_{0}^{2}}< c_{\rm s_{0}}^{2}$ and we have $f^{rr}>0$. As we constrained ourselves to subsonic solutions in this analysis, $\omega^{2}$ is always positive and two real roots exist, thereby ensuring the stability, in qualitative agreement with the results of~\cite{M1980}, derived for the standing wave generated by the perturbation of the velocity potential.


To analyze the traveling waves, we require high frequency waves propagating to large distances. We take trial WKB-like series solution as in the case for Newtonian gravity~(see e.g.~\cite{PSO1980} for details; see also~\cite{M1980} for general relativistic velocity potential perturbation),
\begin{equation}
p_{\omega}(r) = \exp \left[ \sum_{n=-1}^{\infty} \frac{k_{n}(r)}{\omega^{n}} \right].
\label{eqn4.30}
\end{equation}
We substitute this into Eq.~(\ref{eqn4.52}) and set the leading coefficients of individual powers of the frequency $\omega$ to zero. 
It is clear from Eq.s~(\ref{s2}) and (\ref{eqn4.30}) that $k_0$ contributes to the amplitude of $\Psi'$.
Then it turns out that
the leading behaviour of the amplitude of $\Psi'$ in the high frequency approximation we are using is given by
\begin{eqnarray}
\vert \Psi'\vert= \left(\frac{1+g_{rr}v_0^2}{g_{rr}v_0^2c_{\rm s_0}^2}\right)^{\frac14},
\label{ampl}
\end{eqnarray}
which is in principle measurable, since the mass accretion rate is a physical quantity. For analogous expression for the perturbation of the velocity potential, we refer our reader to~\cite{M1980}.

The calculations done so far are valid in general static and spherically symmetric spacetimes. Based on these, we now discuss below two explicit examples $-$
the Schwarzschild and the Rindler spacetime.

\section{Explicit examples}
The metric of the Schwarzschild spacetime can be found from Eq.~(\ref{eqn4.1}), with 
\begin{equation}
g_{tt} = -g_{rr}^{-1} = 1-\frac{2M}{r},~ g_{\theta \theta} = \frac{g_{\phi \phi}}{\sin^2 \theta} = r^{2}.
\label{eqn4.23}
\end{equation}
We shall present the explicit results below, based on our general analysis made in the foregoing sections.

The expressions for the effective acoustic metric and its non relativistic limit can be obtained by substituting for the metric functions (Eq.~(\ref{eqn4.23})) into the general results derived in Section 3. 

As we have discussed in the last section that the standing waves are stable for a general static and spherically symmetric metric, the same holds for the Schwarzschild spacetime also, thereby recovering the result of~\cite{M1980}, derived using the perturbation of the velocity potential. 

Also, using the metric functions into Eq.~(\ref{ampl}), we get
\begin{eqnarray}
\vert \Psi'\vert= \left(\frac{f+v_0^2}{v_0^2c_{\rm s_0}^2}\right)^{\frac14},
\label{ampl2}
\end{eqnarray}
which is the same as that derived earlier in~\cite{Naskar:2007su}, and we have rederived it using our effective acoustic metric. We note that for a spherical flow $v_0$ can never be vanishing. Then since $c_{\rm s_0}\neq0$, the above amplitude can never be divergent, thereby formally establishing the stability of the travelling waves.

It is worthwhile to check the qualitative effect of gravity on the amplitude. We take the radial derivative of the above equation to get
\begin{eqnarray}
\partial_r\vert \Psi'\vert\sim \left(\frac{1}{v_0^2c_{\rm s_0}^2 } \partial_r f -\frac{2\left(1+f\right)}{c_{\rm s_0}^3 } \partial_r c_{\rm s_0}-\frac{2f}{v_0^3}\partial_r v_{0} \right).
\label{ampl3}
\end{eqnarray}
Now, for the spherical flow we are considering both $c_{\rm s_0}$ and $v_0$ should increase monotonically as we move towards the attractor. For the Schwarzschild spacetime we have $\partial_r f=\frac{2}{r^2}$, which is monotonically increasing too, as we move towards the attractor. Putting these all in together, it is clear that the amplitude $\vert \Psi'\vert$ should increase monotonically as we move away from the attractor. We note that this would have been true even we had set $f=1$. The effect of general relativity, which explicitly comes via the factor $f$ in the above equation (and implicitly via $c_{\rm s_0}$  and $v_0$), does not seem to oppose this `shrinking' of the amplitude. This conclusion, keeping in mind our general result of Eq.~(\ref{ampl}), remains valid so long as the metric function $g^{rr}$ increases monotonically with radial distance. Nevertheless, this seems a perfect reasonable condition for the astrophysical black holes.

Here we once again compare our approach with the existing ones. The derivation of the acoustic geometry via perturbation of the velocity potential and the study of the stability analysis were first done in~\cite{M1980}. An energy integral and its time derivative was constructed, and its boundedness was argued. Also, the standing and travelling wave configurations generated by the perturbation of the velocity potential was investigated and the well behaved WKB like high frequency solution were given for the travelling waves. This establishes the stability of the accretion process. Let us now see what we have done here. Firstly, we have seen that 
the propagation of the linear perturbation of the mass accretion rate can be associated with acoustic geometrical structure as well, which seems to be interesting in its own right.
This geometry has similar causal properties like that one gets via the earlier approach. The acoustic geometry we have thus found 
has some mismatch as well in the overall conformal factor, which we have attributed to the non-scalar nature of the mass accretion rate. In fact, a flux is not in general expected to be a scalar, as it depends upon the spatial component of velocity and choice of some spatial 2-surface. In any case, since the mass accretion rate and the velocity potential are two qualitatively different quantities,
our result seems far from obvious. 

We have investigated the standing and travelling wave configurations generated by the perturbation of the mass accretion rate and 
argued the stability of the accretion process. This is in qualitative agreement with the results of~\cite{M1980}. The formalism of~\cite{M1980} establishes strong results on the stability of the accretion process. In particular, this approach seems to be very useful in constructing energy integrals. On the other hand, it seems that if one is interested to understand the behaviour of the mass accretion rate itself, which is also relevant in accretion phenomenon in its own right, our approach seems to provide some convenient manner. In particular,
one can use the general formalism for studying the standing and travelling configurations for the mass accretion rate to numerically investigate their quantitative behaviour.

We also note in particular, an interesting work on stationary state fluid dynamics done in~\cite{michel} for the Schwarzschild spacetime. This estimates the fluid temperature and the energy flux at the critical or sonic point using adiabatic equation of state and boundary conditions at infinity. Generalization for charged fluid is also given. It seems interesting in this context to make a future study of our analysis, for which the accreting fluid is charged and has two components (e.g. ions and electrons). 

We shall also consider another interesting static spacetime, the Rindler spacetime. Since the main focus of this work is accretion astrophysics, we discuss this in the Appendix, to maintain the main course of discussions.

\section{Summary and outlook}

To summarize, the chief agenda of the present paper is the following. 

Firstly, to investigate whether the linear perturbation of the 
mass accretion rate for a radial general relativistic ideal fluid under adiabatic conditions leads to any emergent acoustic geometry,
like that of the perturbation of the velocity potential~\cite{M1980}-\cite{B1999}. If it does, what should be the geometric content
of the theory, keeping in mind that the mass accretion rate (or usually any flux, as it usually associates component of spatial velocity and some choice of spatial 2-surface) manifestly is not a scalar?
Also, could we satisfactorily readdress the stability issues of accretion using this geometry? And finally, to discuss explicit examples using this general formalism.
The physical motivation behind attempting this alternative study comes from the measurable character of the mass accretion rate, and the to study possible wave configurations created by its perturbation. 

The causal structure of the effective acoustic geometry found in Section 3 for a general static and spherically symmetric spacetime is similar as one finds via the perturbation of the velocity potential. We get a conformal factor in Eq.s~(\ref{am1}), (\ref{am2}),
which is not a scalar, and this is the chief mismatch with the velocity potential approach. But this is only expected, due to that
fact that the mass accretion rate is not a scalar, unlike the velocity potential.  Nevertheless, the acoustic causal structures of the two geometries remain the same. Clearly, this result is by no means obvious {\it a priori}, and thus shows the intrinsic physical nature of the acoustic geometry. 

Next we have investigated the wave configurations generated by the perturbation of the mass accretion rate using the acoustic geometry we have found. Solving for the standing and travelling waves, we have argued the stability of the accretion process.
This is in qualitative ageement with the stability associated with the waves generated by the perturbation of the velocity potential~\cite{M1980}.  

Throughout this paper, we have worked with the usual definition of the mass accretion rate~\cite{SJS,FKR}, explicitly based on the symmetry of the spacetime.  
Here we briefly discuss possible directions to extend the current formalism in a more geometric way, which we hope to address in our future studies in details.

Let us first consider the case
of a static and spherically symmetric spacetime~(\ref{eqn4.1}). At the stationary state, where all fluid variables are time independent, it is clear from the continuity equation ($\nabla_a(\rho v^a)=0$) that we may define the mass accretion rate in a covariant way as $\Psi=\int_{S} \rho v_a n^a$, where the integration is done on a 2-sphere (which is preferred in this case, due to the spherical symmetry), and $n^a$ is a unit spacelike normal to it. We may then proceed as earlier and re-do the analysis we have performed. As far as the flow and the perturbation is spherical (which is the case for the present work), this seems to work fine and give the geometry. This is expected, as we have seen in Sec. 3.1 that the acoustic causal structure is necessarily contained in the tensorial part of $f_{\mu\nu}$, so that adopting a covariant approach should not change the main result anyway.     

However, if we wish to include higher modes for the perturbation (like~$Y_{lm}(\theta, \phi)$,~\cite{M1980}), clearly we have to integrate those angular dependence first and we still get an effective $2\times 2$ geometry, instead of the full four dimensional acoustic manifold. One possible way to tackle this issue might be the following. Let us define a modified spherical and stationary state mass accretion rate per unit solid angle as : $\widetilde{\Psi}\sim\rho v_a n^a r^2$, so that when we integrate this over the 2-sphere, we get the usual expression for it. Next, we employ the linear perturbation scheme for $\widetilde{\Psi}$, allow both time and angular depedence for the perturbation, use irrotational conditions and proceed as earlier. Clearly, we may expect this time to derive a full $4\times 4$ geometry.
Afterwords, when we have found explicitly the angular dependence
for the perturbation, we may integrate it to find the full mass accretion rate. 
 We hope to address this issue in details in our future work.

Also, for the case of a non-spherical accretion, the fluid's trajectory creates non-trivial disk geometries. Then, subject to the astrophysical observation, one models those geometries in various ways, e.g.~\cite{FKR, Nowak}. In all these cases, the conventional way to deal with the mass accretion rate is to average over the $\theta=\pi/2$ plane, so that the stationary state mass accretion rate $\Psi_0$ is independedent of $\theta$. Once this has been done, we may proceed as described in the preceeding paragraph to include higher modes in the perturbation of the mass accretion rate.

For cosmological applications, it should be interesting to develop some formalism for the FRW spacetimes. Clearly, due to explicit time dependence of the background geometry, we cannot define a mass accretion rate such as one does in astrophysics.
But a possible way out may be the following. One can define a time dependent mass accretion rate as $\int _{S} \rho v_an^a$, where $S$ is some spatial surface, to be fixed depending upon the specific case we are interested in. Clearly, unlike the astrophysiacl cases, this mass accretion rate will evolve with time. In analogy with the astrophysical case, we make this mass accretion rate, at the zeroth order, to be independent of spatial coordinates. In order to accommodate spatial dependence in the perturbation, we just define a modified mass accretion rate as $\widetilde{\Psi}= \gamma \rho v_an^a $,
where $\gamma$ is the square root of the determinant of the induced metric on the spatial surface, $S$.  
Next we may linearly perturb $\widetilde{\Psi}$ it by taking harmonic spatial dependence $\sim e^{i\vec{k}\cdot \vec{x} }$, and investigate how the perturbation evolves with time.      

For a completely general spacetime without any isometry, it is not so far clear to us how one can define a mass accretion rate, and whether it can still be associated with some internal acoustic manifold. This is in contrast with the velocity potential perturbation~\cite{M1980, unr81}, and we should attribute this to the qualitative and quantitative differences between the velocity poetential and the mass accretion rate. Possibly one may try a $3+1$ foliation of the metric and define a new mass accretion rate using Gauss's theorem as $\int_{S} \rho v_an^a$, where $S$ is some suitable spatial three surface.

However, we note here that as far as the accretion astrophysics is concerned, one is usually intersted in either static spherically symmetric (e.g. the Schwarzschild) or stationary axisymmetric (e.g. the Kerr) spacetimes. In these cases, the formalism we have discussed is expected to work well. Moreover, as one includes non-trivial geometries for the accretion disk in the Schwarzschild or the Kerr spacetimes, the mass accretion rate should show more and more interesting and qualitatively new features to investigate~\cite{FKR, Nowak}. 

One remaining question is, could we just obtain directly $\Psi'$ (\ref{ampl}), from the expression for the perturbed velocity potential? To answer this, we linearly perturb both sides of Eq.~(\ref{g1}). It is clear that in order to determine $\Psi'$, we require the explicit expression for the perturbation of the thermodynamic quantities (such as $\delta \rho$), along with the expression for the perturbation of the velocity potential itself. The expression for $\Psi'$ we have found (Eq.~(\ref{ampl})), on the other hand, is completely determined by the stationary state quantities.

Moreover, we note that in determining $\Psi'$ via some indirect method, we necessarily lose the interesting information that $\Psi'$ is associated with internal acoustic causal structure (Sec. 3.1) and its possible wave configurations (Sec. 3.2).   


Thus if we understand correctly, one possible convenience of using this alternative approach lies in the explicit study of the profile of the mass accretion rate itself. Precisely, apart from the demonstration of the acoustic causal structure associated with the mass accretion rate, which is one of the chief results of this work, we can numerically study the wave configurations (which we have argued to be stable) to get a quantitative estimate. One crucial question now is, does analogous acoustic geometry is associated with axisymmetric flows? For axisymmetric flows
there can be nontrivial disk models for the accretion as well, which as we have mentioned, may bring in qualitative new aspects in the study of the wave configurations. We shall return to this issue soon.

\section*{Appendix A : Fluid mechanics in the Rindler spacetime}
A uniformly accelerating observer in Minkowski spacetime can be described by the Rindler coordinates,   
\begin{equation}
ds^2= -a^{2}x^{2}dt^2+dx^2+dy^2+dz^2,
\label{eqns4.23}
\end{equation}
where the constant
`$a$' represents the uniform acceleration of the observer. The surface `$x=0$' is a Killing horizon, like the black hole horizon and is known as the Rindler horizon. The above spacetime is indeed a flat spacetime, but written in non-inertial coordinates. Apart from being interesting in understanding the near horizon quantum field theory of stationary black holes, the Rindler spacetime can even be relevant in the context of condensed matter systems like a two level atom~\cite{Jin:2014spa}.

We note that even though the Rindler spacetime is flat, unlike the Minkowski spacetime, it is non-inertial due to the observer's acceleration. Hence there are qualitative differences of physics in this frame from the Minkowski spacetime.
The most well known qualitative mismatch is the Unruh effect~\cite{Unruh:1976db}. More recently, investigation of the effect of this non-inertiality on particle decay has been done, see~\cite{Lynch:2014wua} and references therein.
 
We note that in flat inertial spacetime we expect a sonic horizon in presence of an attractor or potential. The Rindler spacetime, although flat, the non-inertiality is expected to take the role of the potential. Precisely, in the frame of the accelerated observer, fluid should be attracted towards the Rindler horizon located at $x=0$.

In this sense, it is interesting to check, if there is any qualitative difference between the fluid mechanics of the Rindler and the Minkowski spacetime, at the classical level.

Let us first consider fluid moving along $x$-direction towards the Rindler horizon. Even though the Rindler spacetime is not spherically symmetric, it is planer symmetric in the $y-z$ plane, in formal analogy with the spherically symmetric spacetimes we considered earlier. We may take the area element orthogonal to the fluid velocity through which the mass accretion rate is measured to be `$dy dz$'. Since the Rindler spacetime has translational isometries along $\partial_y$ and $\partial_z$, we assume that the fluid velocity components and the density does not depend on $y$ and $z$, in an analogous manner of spherical symmetry. Then we define the stationary mass accretion rate, an analogue of Eq.~(\ref{eqn4.9}), $A\rho_0 v_0 x ={\rm const.}$, where $A$ is some area due to integration of $dydz$ element.

Then we study the linear perturbation of the mass accretion rate by making it time dependent.
We may get, using $g_{tt}=a^{2}x^{2}$, $g_{rr} \equiv g_{xx} = 1$ and $\sqrt{-\widetilde{g}}=ax$, 
the effective acoustic metric $f_{\mu\nu}$ and its inverse, from Section 3.  

We shall investigate the behaviour of the fluid trajectory now. We assume that the flow is smooth but otherwise accelerated. 

Before we calculate the expression for $\frac{du_0}{dx}$ for the Rindler metric, we shall evaluate the same for the Schwarzschild black hole spacetime in its near horizon limit. The reason behind this will be clear from the discussions below.

As $r\to 2M$, the Schwarzschild metric can be written as 
\begin{eqnarray}
ds^2\approx -\frac{x^2}{16M^2}dt^2+dx^2+4M^2d\Omega^2,
\label{r1}
\end{eqnarray}
where we have defined $\sqrt{2M(r-2M)}=\frac{x}{2}$. The `$t-r$' part now looks alike Eq.~(\ref{eqns4.23}) with the identification : $a^2=\frac{1}{16M^2}$~\footnote{For a black hole with `large enough' horizon size, we can replace the two sphere by an Euclidean two plane, and then the metric (\ref{r1}) looks entirely alike the Rindler. However, this is not going to alter our result anyway.}.

Let us now compute the expression for $\frac{du_0}{dx}$ for the Schwarzschild metric in the near horizon limit. Using Eq.s~(\ref{r1}) and (\ref{eqs3}), we find
\begin{equation}
\frac{du_0}{dx} = \frac{u_{0}\left(1-u_{0}^{2}\right)\left(c_{\rm s_{0}}^{2}-1\right)}{\left(u_{0}^{2}-c_{\rm s_{0}}^{2}\right)x}.
\label{r2}
\end{equation}
Using Eq.s~(\ref{eqs1}) and the normalization condition $v_{\mu}v^{\mu}=-1$, it is easy to find that $u_0\to 1$ in the near horizon limit, which can be understood as the fact that the black hole horizon is a null hypersurface. Now, between ultrarelativistic and non-relativistic limits the adiabatic constant $\gamma$ appearing in the equation of state ranges from $\frac43 \leq \gamma\leq \frac53$~(see, e.g.~\cite{weinberg}), respectively. Then the positivity of the fluid's energy (Eq.~(\ref{eqn4.10})) ensures that $c_{\rm s_{0}}<1$, always. 

Then it is clear that we can never have a sonic horizon on the black hole event horizon, $x=0$, because otherwise from Eq.~(\ref{r2}) we must have $c_{\rm s_{0}}^2=1$ there. Using L'Hospital's rule, we also find 
\begin{equation}
\lim_{x\to 0}\frac{du_0}{dx} =0,
\label{r3}
\end{equation}
which can be understood as an outcome of the smoothness of the near horizon geometry.

Let us now turn to the actual Rindler spacetime, Eq.~(\ref{eqns4.23}). The expression for $\frac{du_0}{dx}$ looks exactly the same (Eq.~(\ref{r2})) in this case also,
but the difference is, we have now all possible values of $x$. Then it is clear that if we have a sonic horizon, $u_{0}^{2}-c_{\rm s_{0}}^{2}=0$ for $x\neq 0$, we must have $u_{0}^{2}=1=c_{\rm s_{0}}^{2}$ there. But we have argued that $u_{0}$ can reach the speed of light only when $x\to 0$, and $c_{\rm s_{0}}<1$ always. Putting these all in together we conclude, there can be no sonic horizon in the Rindler spacetime for smooth flow. Clearly, on the Rindler horizon Eq.~(\ref{r3}) holds as well.

Possibly, this result indicates the fact that, as far as mechanics or thermodynamics is concerned at the classical level, there should be no difference 
of physical phenomenon as far as mere coordinate transformations are concerned. In other words, the absence of the sonic point for smooth flows in the Rindler spacetime, even though it is non-inertial, should be attributed to the absence of sonic points in the Minkowski spacetime. It should be interesting to investigate this effect at the quantum level.

This result can easily be generalized for flow along all four directions as the following. Since we have translational isometries along $\partial_y$ and $\partial_z$, we can define conserved specific momenta
along those directions : $\displaystyle k_y=-\frac{v_y}{v_t}$ and $\displaystyle k_z=-\frac{v_z}{v_t}$. This is analogous to defining specific angular momentum for axisymmetric spacetimes.
 
The normalization condition for the fluid's four-velocity becomes
\begin{eqnarray}
a^2x^2(v^t)^2\left[1-a^2x^2k^2\right]=1+v^2,
\label{r5}
\end{eqnarray}
where we have written $k^2=k_y^2+k_z^2$. Next we define the transformations for a local frame
\begin{eqnarray}
v_0^t&=&\frac{1}{\sqrt{a^2x^2\left(1-a^2x^2k^2\right) }}\frac{1}{\sqrt{1-u_0^2}},\quad
v_0=\frac{u_0}{ \sqrt{1-u_0^2}},\nonumber\\
v^y_0&=&\frac{axk_y}{\sqrt{\left(1-a^2x^2k^2\right) }}\frac{1}{\sqrt{1-u_0^2}},\quad
v^z_0=\frac{axk_z}{\sqrt{\left(1-a^2x^2k^2\right) }}\frac{1}{\sqrt{1-u_0^2}}.
\label{r6}
\end{eqnarray}
Then follwoing methods of Scetion 2, we find after a straightforward calculation a generalization of Eq.~(\ref{r2}),
\begin{eqnarray}
\frac{du_0}{dx}= \frac{u_0(1-u_0^2)}{x(u_0^2-c_{\rm s_0}^2)}\left[c_{\rm s_0}^2-\frac{1}{1-k^2a^2x^2}\right].
\label{r7}
\end{eqnarray}
The normalization condition, Eq.~(\ref{r5}), shows that we have always $a^2x^2k^2<1$. Then similar argument as earlier proves that there can be no sonic point, $u_0^2=c_{\rm s_0}^2$.

\vskip 1cm
\section*{Acknowledgement}
Long term visit of D. B Ananda at Harish-Chandra Research Institute (HRI, India) has been supported by the planned project fund of the
Cosmology and the High Energy Astrophysics subproject of HRI. Majority of S. Bhattacharya's work was done when he was a post doctoral fellow in HRI, India. His current research is implemented under the ``ARISTEIA II" Action of the Operational Program ``Education and Lifelong Learning" and is co-funded by the European Social Fund (ESF) and Greek National Resources.
He thanks Jayanta K Bhattacharjee for useful discussions and encouragement. 
The authors sincerely acknowledge anonymous referee for pointing out some errors in calculation and for many valuable suggestions.


\end{document}